\documentclass[lettersize,journal]{IEEEtran}
\usepackage{amsmath,amsthm,amsfonts,amssymb,amscd}
\usepackage[caption=false,font=normalsize,labelfont=sf,textfont=sf]{subfig}
\usepackage{multirow,booktabs}
\usepackage[table]{xcolor}
\usepackage{lastpage}
\usepackage{enumitem}
\usepackage{fancyhdr}
\usepackage{mathrsfs}
\usepackage{wrapfig}
\usepackage{setspace}
\usepackage{multicol}
\usepackage{cancel}
\usepackage{amsmath}
\usepackage{graphicx} 
\usepackage{float} 
\usepackage{caption2}
\usepackage{algorithm}
\usepackage{algpseudocode} 
\ifCLASSINFOpdf
\allowdisplaybreaks[4]
\else
\fi
\newtheorem{theorem}{\itshape Theorem}

\newtheorem{problem}{\itshape Problem}

\begin{document}
	\setlength{\parskip}{0pt}
	\title{Privacy Preservation by Intermittent Transmission in Cooperative LQG Control Systems}
	\author{Wenhao Lin$^\star$, Yuqing Ni$^\dag$, Wen Yang$^\star$, and Chao Yang$^\star$
		\thanks{$^\star$: Key Laboratory of Smart Manufacturing in Energy Chemical Process, Ministry of Education, Dept. of Automation, East China University of Science and Technology, Shanghai, China, 200237.
			Email: y30220957@mail.ecust.edu.cn, \{weny, yangchao\}@ecust.edu.cn.
			$\dag$: Key Laboratory of Advanced Process Control for Light Industry (Ministry of Education), School of Internet of Things Engineering, Jiangnan University, Wuxi, China.
			Email: yuqingni@jiangnan.edu.cn.
			The corresponding author is Chao Yang.} 
	}
	\maketitle
	
	
	\IEEEpeerreviewmaketitle
	
	
	\begin{abstract}
		In this paper, we study a cooperative linear quadratic Gaussian (LQG) control system with a single user and a server. In this system, the user runs a process and employs the server to meet the needs of computation. However, the user regards its state trajectories as privacy. Therefore, we propose a privacy scheme, in which the user sends data to the server intermittently. By this scheme, the server's received information of the user is reduced, and consequently the user's privacy is preserved. In this paper, we consider a periodic transmission scheme. We analyze the performance of privacy preservation and LQG control of different transmission periods. Under the given threshold of the control performance loss, a trade-off optimization problem is proposed. Finally, we give the solution to the optimization problem.
	\end{abstract}
	
	
	\section{Introduction}
	Networked control systems (NCSs) have been widely deployed in many fields of our lives for a long time \cite{hespanha2007survey}, such as smart grids, intelligent transportation and intelligent manufacturing. Due to the reliability of information and communication technologies, NCSs can effectively perform a multitude of complex functions. In NCSs, various components of the system are connected through a network, which enables them to cooperate more conveniently. For example, the system could belong to different parties, and the system's operation is achieved through the cooperative efforts by various parties. We call systems characterized by this behavior as \textit{cooperative networked control systems}. Such cooperative behavior should indeed become a prevailing trend in the future. For example, in the domain of computer science, data owners frequently outsource their data to servers to access functions such as storage, management, and computation, thus significantly reducing the cost associated with data maintenance \cite{mendhe2012survey}. For example, cloud services and cooperative networked control systems serve as robust enablers for this paradigm, offering robust support and facilitation. \par 
	In this paper, we consider a basic cooperative control system called \textit{user-server system}, which is composed of a single user and a server. In this system, the user needs to ask the server for services and thus shares information with it, but the information may contain what the user think is privacy. Meanwhile, the server is considered to be ``honest but curious", i.e., it will honestly provide the user with services it needs, but also analyzes the user's information and explores its behavior as much as possible. Therefore, it is necessary to study the privacy preservation at the user side in cooperative NCSs, which is the focus of this paper. \par 
	\subsection{Related Studies}
	Several various frameworks are proposed to study privacy preservation based on different understandings. Three important frameworks are \textit{differential privacy} \cite{le2013differentially}, \textit{information-theoretic} \cite{nekouei2019information}, and \textit{homomorphic encryption} \cite{yi2014homomorphic}.\par 
	Differential privacy, which is originally applied in the privacy preservation of static databases \cite{dwork2006differential}, due to its strong privacy guarantees, has attracted attention recently. The Laplace mechanism for differential privacy is proposed in \cite{dwork2006calibrating}. Yu Kawano et al. \cite{2020A} analyzed the differential privacy level of the Laplace mechanism in the cloud-based control system. Jerome Le Ny et al. \cite{le2017differentially} developed approximate MIMO filters to implement differential privacy guarantees. Kasra Yazdani et al. \cite{Yazdani2023Differentially} presented a multi-agent LQ control framework which guarantees differential privacy. Kwassi H. Degue et al. \cite{Degue2023Cooperative} studied a cooperative differentially private LQG control problem with measurement aggregation,  and proposed a two-stage architecture to solve it. Martin Abadi et al. \cite{abadi2016deep} studied the applications of differential privacy in deep learning and developed new algorithmic techniques for it. Arik Friedman et al. \cite{friedman2010data} applied differential privacy in data mining and proposed an improved algorithm. Tie Ding et al. \cite{Ding2023Differentially} proposed a algorithm that guarantees differential privacy to deal with a constrained resource allocation problem. Francois Gauthier et al. \cite{Gauthier2023Personalized} proposed a privacy-preserving personalized graph federated learning algorithm by applying differential privacy.  \par 
	Information-theoretic framework focuses on the nature of information, using quantitative metrics such as conditional entropy, mutual information, and directed information to evaluate privacy level. Peng Hao et al. \cite{Hao2019Integrating} applied the proposed offline maximum entropy-based quantization rule to the security in Internet of things.  Ehsan Nekouei et al. \cite{nekouei2018privacye} designed a privacy-aware estimator by an entropy constrained approach. Ruoxi Jia et al. \cite{jia2017privacy} used mutual information between the location trace and the reported occupancy measurement as a privacy metric, and designed a scheme that can balance the privacy and control performance. Ehsan Nekouei et al. \cite{nekouei2018privacy} considered a multi-sensor estimation problem, where the conditional entropy of each sensor is applied as a privacy level. Causal conditioned directed information was proposed in \cite{kramer2003capacity}. Takashi Tanaka et al. \cite{tanaka2017directed} studied an optimization problem of cloud-based LQG systems, in which Kramer's causal conditioned directed information is used as a privacy metric, and an algorithm is proposed to solve the problem. \par  
	Homomorphic encryption permits third parties to operate on the encrypted data without requiring prior decryption. Kiminao Kogiso et al. \cite{kogiso2015cyber} proposed a controller encryption scheme by using the modified homomorphic encryption.  Farhad Farokhi et al. in \cite{alexandru2019encrypted} applied labeled homomorphic encryption in LQG control systems. Farhad Farokhi et al. \cite{farokhi2016secure} considered the sensors using the Paillier encryption which is a semi-homomorphic encryption in cloud-based systems. Mohammad Faiz et al. \cite{faiz2022improved} used the particle swarm optimization to improve homomorphic encryption for cloud security. 
	
	\subsection{The Study of This Paper}
	In this paper, we consider a basic user-server cooperative networked LQG control system. The user shares its information with the server and asks the server to compute control inputs. From the user’s view, certain shared information is private. To guarantee the privacy level, we consider designing a privacy scheme at the user side.   
	The main contributions of this paper could be summarized as follows. 
	\begin{enumerate}
		\item We design a novel privacy preservation scheme, which is achieved by intermittent transmission, for the user in a user-server cooperative networked control system, which is based on a closed-loop LQG control system.
		\item For our proposed intermittent transmission scheme, we analyze the estimation performance for infinite-time horizon at the estimator, and propose a novel privacy metric.
		\item We analyze the privacy level and the LQG control performance, and study the trade-off problems between them.
	\end{enumerate}
	\textit{Notations:} $\mathbb{Z}_{+}$ is the set of non-negative integers and $k \in \mathbb{Z}_{+}$
	is the time index. $\mathbb{N}$ is the set of natural numbers. $\mathbb{R}$ is the set of real numbers. $\mathbb{R}^{n}$ is $n$-dimensional Euclidean space. $\mathbb{S}^{n}_{+}$ (and $\mathbb{S}^{n}_{++}$) is the set of $n$ by $n$ positive semi-definite matrices (and positive definite matrices); when $X \in \mathbb{S}^{n}_{+} $ (and $\mathbb{S}^{n}_{++}$), it is written as $X \geq 0$ (and $X>0$). It is defined by $X \geq Y$ if $X-Y \in \mathbb{S}^{n}_{+}$. $\mathbf{E}(\cdot)$ or $\mathbf{E}[\cdot]$ is the expectation of a random variable and $\mathbf{E}(\cdot|\cdot)$ or $\mathbf{E}[\cdot|\cdot]$ is the conditional expectation. $tr(\cdot)$ is the trace of a matrix. $\rho(\cdot)$ is the spectrum radius of a matrix.
	
	\section{Problem Setup}
	In this section, we introduce the model of the ordinary user-server system for cooperative LQG control, and propose a framework that simultaneously balances the control performance and the user's privacy preservation.
	
	\subsection{Ordinary System Model }
\begin{figure}[htbp!]
		\centering
		\includegraphics[scale=0.29]{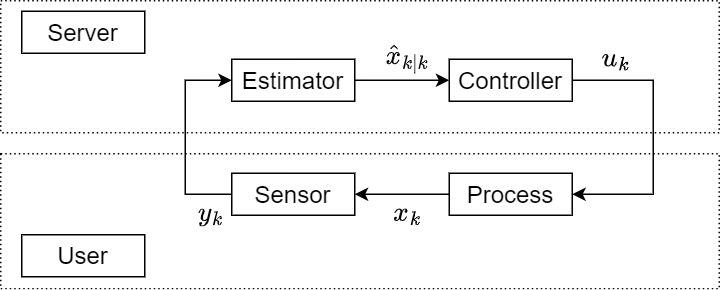}
		\caption{The ordinary cooperative LQG control system.}
		\label{Ordinarymodel}
	\end{figure}

	The ordinary user-server system for cooperative LQG control is illustrated in Fig. \ref{Ordinarymodel}. The user has a state process to control and a sensor which measures the process states. The server has sufficient computation capability. The user sends its data to the server for the optimal LQG control input.
	 \par 
	The user needs to control the following dynamic process:
	\begin{align}
		x_{k+1}=Ax_{k}+Bu_{k}+w_{k},
	\end{align}	
	where $x_{k}\in \mathbb{R}^{n}$ is the state of the process at time $k$, $u_{k}\in \mathbb{R}^{m}$ is the control input, and $w_{k}$ is the Gaussian white noise whose mean is zero and covariance is $Q (Q\geq0)$. The matrix $A \in \mathbb{R}^{n \times n}$ is the system matrix, and $B \in \mathbb{R}^{n \times m}$ is the input matrix. The initial condition is assumed that $x_{0}$ is Gaussian with distribution $\mathcal{N}\left(\bar{x}_{0},\Sigma_{0}\right)$. The time horizon of the process is assumed to be infinite. We assume that $(A,B)$ is controllable.\par 
	The user cannot directly get the state value and has a sensor to measure the state as follows:
	\begin{align}
		y_{k}=Cx_{k}+v_{k},
	\end{align} 
	where $y_{k} \in \mathbb{R}^{q}$ is the measurement of $x_k$, and $v_k$ is the Gaussian white noise whose mean is zero and covariance is $R (R>0)$. The matrix $C \in \mathbb{R}^{q \times n}$ is the measurement matrix. The noises $\left\{w_{k}\right\}$ and $\left\{v_{k}\right\}$ and the initial state $x_{0}$ are mutually independent of each other. The pairs $(A,\sqrt{Q})$ and $(C,A)$ are assumed to be stabilizable and detectable, respectively.\par 
	For the user, its demand is the optimal LQG control input. The considered infinite-time quadratic objective function for the LQG control is denoted as $\mathcal{J}$, and is defined as follows:
	\begin{align}
		\mathcal{J}\triangleq \lim_{N\to \infty} \frac{1}{N} \mathbf{E}\Big[\sum_{k=0}^{N-1}(x_{k}'Wx_{k}+u_{k}'Uu_{k})\Big],
	\end{align}
	where $W$ and $U$ are weight matrices satisfying $W \ge 0$ and $U > 0$, and the expectation is taken with respect to the possible randomness.\par 
	In this cooperative system, the user employs the server for the needs of control, and the server computes the optimal control input and returns it to the user. The server serves as the estimator and the controller. Since the user's process state is unavailable, it needs to be estimated by the server at first, and the server computes the optimal control input based on it. The system parameters $A, B, W, U$ for LQG control, $C, Q, R$ for state estimation, and the initial condition $x_{0}$ with Gaussian distribution $\mathcal{N}(\bar{x}_{0},\Sigma_{0})$ are shared by the user with the server, and the user is also supposed to provide $y_{k}$ at each time $k$. \par 
	When the process begins, the user sends $y_{k}$ to the server at each time $k$. Let $Y_{k}\triangleq\left\{y_{1},y_{2},\dots,y_{k}\right\}$ denote the set of the received measurements. Then the server computes the \textit{a priori} and \textit{a posteriori} estimates $\hat{x}_{k|k-1}$ and $\hat{x}_{k|k}$ defined as follows:
	\begin{align*}
		\hat{x}_{k|k-1}&\triangleq \mathbf{E}[x_{k}|Y_{k-1}],\\
		\hat{x}_{k|k}&\triangleq \mathbf{E}[x_{k}|Y_{k}].
	\end{align*}
	Meanwhile, we define $P_{k|k-1}$ and $P_{k|k}$ as the estimate error covariance matrices associated with $\hat{x}_{k|k-1}$ and $\hat{x}_{k|k}$, respectively:
	\begin{align*}
		P_{k|k-1}&\triangleq \mathbf{E}\Big[(x_{k}-\hat{x}_{k|k-1})(x_{k}-\hat{x}_{k|k-1})'|Y_{k-1}\Big],\\
		P_{k|k}&\triangleq \mathbf{E}\Big[(x_{k}-\hat{x}_{k|k})(x_{k}-\hat{x}_{k|k})'|Y_{k}\Big].
	\end{align*}
	The server runs the standard Kalman filter under the given initial condition $\hat{x}_{0|0}=\bar{x}_0$ and $P_{0|0}=\Sigma_{0}$, and the estimates and the associated error covariances satisfy the following recursion:
	\begin{align}
		\hat{x}_{k|k-1}&=A\hat{x}_{k-1|k-1}+Bu_{k-1},\\
		P_{k|k-1}&=AP_{k-1|k-1}A'+Q,\\
		K_{k}&=P_{k|k-1}C'(CP_{k|k-1}C'+R)^{-1},\\
		\hat{x}_{k|k}&=\hat{x}_{k|k-1}+K_{k}(y_{k}-C\hat{x}_{k|k-1}),\\
		P_{k|k}&=(I-K_{k}C)P_{k|k-1}.
	\end{align} 
	Then the server computes the optimal control input and returns it to the user. The computation follows
	\begin{align}
		S_{T}=\; &W,\\
		S_{k}=\; &A'S_{k+1}A+W \nonumber \\
		&-A'S_{k+1}B(B'S_{k+1}B+U)^{-1}B'S_{k+1}A,  \\ 
		L_{k}=\; &-(B'S_{k+1}B+U)^{-1}B'S_{k+1}A, \\
		u_{k}^{*}=\; &L_{k}\hat{x}_{k|k},
	\end{align}
	where eqn. (9)-(11) are computed before the process. \par 
	For the Kalman filter used in this paper, it has an asymptotic steady error covariance \cite{anderson2012optimal}, and we denote it as
	\begin{align*}
		\lim_{k\to \infty}P_{k|k}=\bar{P}.
	\end{align*}
	The error covariance converges at an exponential rate. Hence, for convenience, under the considered infinite-time horizon, we assume that the initial covariance $P_{0|0}=\bar{P}$. Consequently for all $k \geq 1$, we have
	\begin{align*}
		P_{k|k}&=\bar{P}, \\ 
		K_{k}&=K.
	\end{align*} 
	Meanwhile, for the LQG controller, the matrix $S_{k}$ also converges to a steady value and becomes time-invariant, and we have
	\begin{align*}
		\lim_{N\to \infty}S_{k}=S, 
	\end{align*}
	where 
	\begin{align*}
		S=A'SA+W-A'SB(B'SB+U)^{-1}B'SA.
	\end{align*}
	Then we evaluate the performance of the optimal LQG control. The optimal objective function value in infinite-time horizon, denoted as $\mathcal{J}^{*}$, is given as follows \cite{bertsekas2012dynamic}:
	\begin{align}
		\mathcal{J}^{*}= tr(SQ)+tr(\Phi \bar{P}),
	\end{align}
	where 
	\begin{align*}
		\Phi=A'SB(B'SB+U)^{-1}B'SA.
	\end{align*}
	
	\subsection{Privacy Preservation Model}
	\begin{figure}[htbp!]
		\centering
		\includegraphics[scale=0.26]{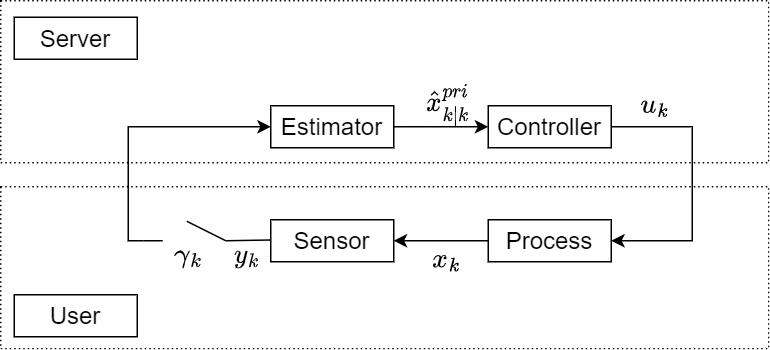}
		\caption{The privacy scheme is used.}
		\label{Privacymodel}
	\end{figure}
	In the ordinary cooperative LQG control system, the user's  information of state is also known to the server. However, the user treats the information of state as its privacy, so it decides to employ a privacy scheme to preserve the privacy. \par 
	In this paper, we consider applying a scheme of intermittent transmission (Fig. \ref{Privacymodel}). Let $\gamma_{k}$ be the user's decision variable to control the transmission of the measurement $y_{k}$ to the server at time $k$, i.e., if $\gamma_{k}=1$, $y_{k}$ is sent to the server, and if $\gamma_{k}=0$, $y_{k}$ is not sent. Intuitively, the user transmits measurements intermittently, which reduces the amount of information obtained by the server, thus preserving the privacy. \par 
	We consider the scenario that the intervals between measurement transmissions are identical, i.e., the transmission is periodic. We assume at time $k=1$ the user chooses to transmit the measurement, and denote the transmission period as $T$. For $l \in \mathbb{Z}_{+}$, the scheme can be formulated as follows:
	\begin{align}
		\gamma_{k}=\begin{cases}
			1, & k=lT+1, \\
			0, & k \neq lT+1. 
		\end{cases} 
	\end{align}  
	An illustration of $T=3$ is shown as follows:
	\begin{align*}
		\underbrace{100}_{T=3}100\; 100\; 1\dots 
	\end{align*}
	
	\subsection{Privacy Metric}
	In the system under the privacy scheme, the server is unable to receive all the measurements from the user. Let $\tilde{Y}_{k} \triangleq \left\{\gamma_{1}y_{1}, \gamma_{2}y_{2}, \dots, \gamma_{k}y_{k}\right\}$ denote the set of measurements actually received by the server. The \textit{a priori} and \textit{a posteriori} estimates in this scenario, denoted as $\hat{x}_{k|k-1}^{pri}$ and $\hat{x}_{k|k}^{pri}$, defined as follows:
	\begin{align*}
		\hat{x}_{k|k-1}^{pri}& \triangleq \mathbf{E}[x_{k}|\tilde{Y}_{k-1}], \\
		\hat{x}_{k|k}^{pri}& \triangleq \mathbf{E}[x_{k}|\tilde{Y}_{k}].
	\end{align*}
	Meanwhile, the estimate error covariance matrices associated with $\hat{x}_{k|k-1}^{pri}$ and $\hat{x}_{k|k}^{pri}$ are denoted as $P_{k|k-1}^{pri}$ and $P_{k|k}^{pri}$, defined as follows:
	\begin{align*}
		P_{k|k-1}^{pri}& \triangleq \mathbf{E}\Big[(x_{k}-\hat{x}_{k|k-1}^{pri})(x_{k}-\hat{x}_{k|k-1}^{pri})'|\tilde{Y}_{k-1}\Big], \\
		P_{k|k}^{pri}& \triangleq \mathbf{E}\Big[(x_{k}-\hat{x}_{k|k}^{pri})(x_{k}-\hat{x}_{k|k}^{pri})'|\tilde{Y}_{k}\Big].
	\end{align*}
	When $\gamma_{k}=1$, they evolve as 
	\begin{align}
		\hat{x}_{k|k-1}^{pri}&=A\hat{x}_{k-1|k-1}^{pri}+Bu_{k-1},\\
		P_{k|k-1}^{pri}&=AP_{k-1|k-1}^{pri}A'+Q,\\
		K_{k}^{pri}&=P_{k|k-1}^{pri}C'(CP_{k|k-1}^{pri}C'+R)^{-1},\\
		\hat{x}_{k|k}^{pri}&=\hat{x}_{k|k-1}^{pri}+K_{k}^{pri}(y_{k}-C\hat{x}_{k|k-1}^{pri}),\\
		P_{k|k}^{pri}&=(I-K_{k}^{pri}C)P_{k|k-1}^{pri},
	\end{align} 
	and when $\gamma_{k}=0$, we have
	\begin{align}
		\hat{x}_{k|k}^{pri}&=\hat{x}_{k|k-1}^{pri}, \\
		P_{k|k}^{pri}&=P_{k|k-1}^{pri}.
	\end{align}
	Hence, the privacy scheme will cause the increase of the estimate error covariance. Based on it, we propose the privacy metric to measure the quality of privacy preservation. For each time $k$, we define the privacy metric as the deviation between the estimate error covariance after using the privacy scheme and the one in the ordinary system. Specifically, it is defined as follows:
	\begin{align*}
		\mathcal{Q}_{privacy}^{k}\triangleq P_{k|k}^{pri}-P_{k|k}.
	\end{align*} 
	To study the privacy performance of infinite-time horizon, we define the averaged privacy metric at all time as 
	\begin{align*}
		\mathcal{Q}_{privacy}=\lim_{N\to \infty} \frac{1}{N}\sum_{k=0}^{N-1}\mathcal{Q}_{privacy}^{k}.
	\end{align*}
	\subsection{Problems to Study}
	With respect to the proposed privacy preservation scheme, our attention will be directed towards the following problems.
	\begin{itemize}
		\item[1] 
		We evaluate the privacy preservation effectiveness. Meanwhile, the privacy scheme incurs a decrease in service quality, which results in a relative loss for LQG control performance, denoted by $\mathcal{Q}_{LQG}$. We also analyze this performance degradation.
		\item[2] 
		We study the trade-off between the privacy preservation and LQG
		control performances:
		\begin{align*}
			\text{max}\quad\quad &\text{tr}(\mathcal{Q}_{privacy})\\
			\text{s.t.}\quad\quad &\mathcal{Q}_{LQG}\leq\alpha,
		\end{align*}
		where $\alpha>0$ represents a given loss level of LQG control performance. 
	\end{itemize}
	
	\section{Performance Analysis}
	In this section, we firstly introduce preliminaries of Kalman filtering and present the explicit forms of privacy and LQG control performances. Secondly, we study and solve the trade-off optimization problem about them. 
	
	\subsection{Kalman Filtering Preliminaries}
	Before stating the main results of the paper, we provide a summary of necessary properties of the Kalman filter. \par 
	The function $h(X):\mathbb{S}_{+}^{n}\rightarrow \mathbb{S}_{+}^{n}$ is defined as
	\begin{align}
		h(X)\triangleq AXA'+Q,
	\end{align}
	which is also called the Lyapunov function. The function $\tilde{g}:\mathbb{S}_{+}^{n}\rightarrow \mathbb{S}_{+}^{n}$ is defined as follows: 
	\begin{align}
		\tilde{g}(X)\triangleq X-XC'(CXC'+R)^{-1}CX.
	\end{align}
	Define $g=\tilde{g}\circ h(X)$. In the ordinary system, we have
	\begin{align}
		P_{k|k-1}&=h(P_{k-1|k-1}), \\
		P_{k|k}&=g(P_{k-1|k-1}).
	\end{align}
	When $k \to \infty$, it follows
	$\bar{P}=g(\bar{P})$, and $\bar{P}$ is the unique solution to $X=g(X)$.
	
	\subsection{Privacy Performance}
	For the estimate error covariance at the server side, we have the following results.
	\begin{theorem}
		Given the transmission period $T$, based on the assumption that $(C,A^{T})$ is detectable, we have
		\begin{align*}
			\lim_{l\to \infty}P_{lT+1|lT+1}^{pri}=\tilde{P},  
		\end{align*}
		in which $\tilde{P}$ is the unique solution to the following equation:
		\begin{align*}
			\tilde{P}=g \circ h^{T-1}(\tilde{P}).
		\end{align*}
		Meanwhile two limits exist as follows:
		\begin{align*}
			\liminf_{k\to\infty}P_{k|k}^{pri}&=\tilde{P},  \\
			\limsup_{k\to\infty}P_{k|k}^{pri}&=h^{T-1}(\tilde{P}).
		\end{align*}
	\end{theorem} 
	\begin{IEEEproof}
		In appendix.
	\end{IEEEproof} 
	From Theorem 1, $\tilde{P}$ is the error covariance at the time when the measurement is transmitted, and the server's calculation of $P_{k|k}^{pri}$ follows eqn. (21) until the next transmission. Hence, we can see that the estimate error covariance at the server side is
	\begin{align}
		P_{k|k}^{pri}=\begin{cases}\tilde{P},&\textit{if} \enspace \gamma_k=1,\\
			h(P_{k-1|k-1}^{pri}),&\textit{if} \enspace \gamma_k=0.\end{cases}
	\end{align}
	In the original system, we have $P_{k|k}=\bar{P}$. We can now obtain the metric of privacy performance:
	\begin{align}
		\mathcal{Q}_{privacy}=\frac{1}{T}\sum_{i=0}^{T-1}h^{i}(\tilde{P})-\bar{P}.
	\end{align}

	\subsection{LQG Control Performance}
	The server computes the control policy based on the user's state estimate. Adding the privacy scheme, the control input computed by the server is
	\begin{align}
		u_{k}=L_{k}\hat{x}_{k|k}^{pri}.
	\end{align}
	From the above equation, we can see that a deviation is caused by the privacy scheme for the control input. This deviation is caused by the estimate without transmitting the measurement. At that time, the result corresponds to the \textit{a prior} estimate of the original system, which makes the control law $\left\{u_{k}\right\}$ be non-optimal under the scheme. Hence, the resulting state trajectories $\left\{x_{k}\right\}$ is also non-optimal. Therefore, privacy scheme can enhance the privacy performance while it will lead to the sacrifice in the LQG performance. The optimal LQG performance under the privacy scheme, denoted as $\mathcal{O}^{*}$, is studied below.
	\begin{theorem}
		Under the privacy scheme, it holds that
		\begin{align*}
			\mathcal{O}^{*}= tr(SQ)+tr\big(\Phi \frac{1}{T} \sum_{i=0}^{T-1} h^{i}(\tilde{P})\big),			
		\end{align*}
		where
		\begin{align*}
			\Phi=A'SB(U+B^{'}SB)^{-1}B'SA.
		\end{align*}
	\end{theorem}
	\begin{IEEEproof}
		In appendix.
	\end{IEEEproof}
	 Now we can define the quality loss in LQG performance:
	\begin{align*}
		\mathcal{Q}_{LQG}&\triangleq \mathcal{O}^{*}-\mathcal{J}^{*} 
		=tr(\Phi \mathcal{Q}_{privacy}).
	\end{align*}	

	\subsection{Optimization Problem}
	Through previous study, we can see that privacy preservation is gained while requiring the sacrifice in LQG control performance. Hence, an optimization problem is proposed to study the trade-off: to maximize the privacy metric $\mathcal{Q}_{privacy}$ while the loss of LQG performance $\mathcal{Q}_{LQG}$ is under a given level $\alpha$. The problem is as follows:
	\begin{problem}
		\begin{align*}
			\max_{T}\quad\quad &tr(\mathcal{Q}_{privacy})\\
			\mathrm{s.t.}\quad\quad &tr(\Phi \mathcal{Q}_{privacy}) \leq \alpha, \\
			& \mathcal{Q}_{privacy}=\frac{1}{T}\sum_{i=0}^{T-1}h^{i}(\tilde{P})-\bar{P}.
		\end{align*}
	\end{problem}
	The function involved in this problem is discrete, so we consider using dichotomy (Algorithm \ref{Algorithm1}) to solve it.
	\begin{algorithm}[h]  
		\caption{Optimal $T$ by Using Dichotomy}  
		\label{Algorithm1}  
		\begin{algorithmic}[1]  
			\footnotesize {
			\Require  
			$\mathcal{Q}_{LQG}(T)$: the objective function;  
			$T_{l}$: the left bound of search range;
			$T_{r}$: the right bound of search range;
			$\alpha$: the given threshold;
			\Ensure  
			$T^{*}$: optimal $T$;  
			\State Initialize $T^{*}$;
			\Repeat  
			\State compute the mid index and round down: $T_{m}=floor((T_{l}+T_{r}/2))$;
			\State compute the objective value of mid index: $\mathcal{Q}_{m}=\mathcal{Q}_{LQG}(T_{m})$;
			\If{$\mathcal{Q}_{m}<\alpha$}
			\State $T^{*}=T_{m}$;
			\State $T_{l}=T_{m}+1$;
			\Else 
			\State $T_{r}=T_{m}-1$;
			\EndIf
			\Until{$T_{l}>T_{r}$} }
		\end{algorithmic}  
	\end{algorithm}  
	\section{Example}                                      
\begin{figure}[htbp!]
		\centering
		\includegraphics[scale=0.33]{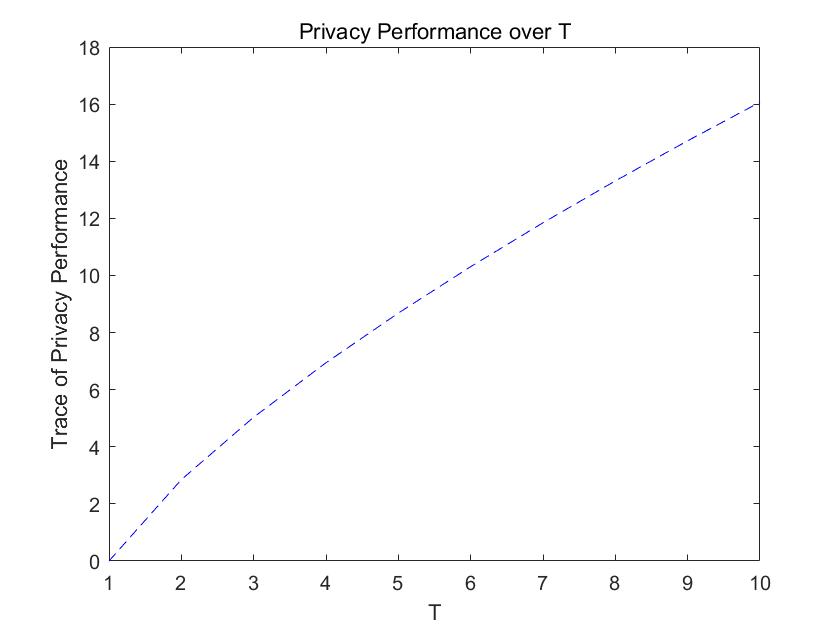}
		\caption{$\mathcal{Q}_{privacy}$ with $T$ varying from 1 to 10.}
		\label{Privacyperformance} 
	\end{figure}

	We consider a second-order system with infinite-time horizon, whose parameters are as follows:
	\begin{align*}
		&A=\begin{bmatrix}
			0.19 & 0.46 \\
			0.31 & 0.8
		\end{bmatrix}, \quad
		B=\begin{bmatrix}
			2 \\ 
			1
		\end{bmatrix}, \\
		&C=\begin{bmatrix}
			1 & 0
		\end{bmatrix}, \quad
		Q=\begin{bmatrix}
			1.9 & 0.9 \\
			0.9 & 2.8
		\end{bmatrix}, \\
		&R=1, \quad
		W=\begin{bmatrix}
			1.5 & 0.5 \\
			0.5 & 1.5
		\end{bmatrix}, \quad U=1.
	\end{align*}
	Firstly we plot the traces of $\mathcal{Q}_{privacy}$ in different $T$ (Fig. \ref{Privacyperformance}).
	 \par 
	We find the curve is monotonically increasing, and it means the larger $T$ is, the greater the period between transmitting the measurement is, and the better the privacy performance is, which conforms to eqn. (27). \par
\begin{figure}[htbp!]
		\centering
		\includegraphics[scale=0.33]{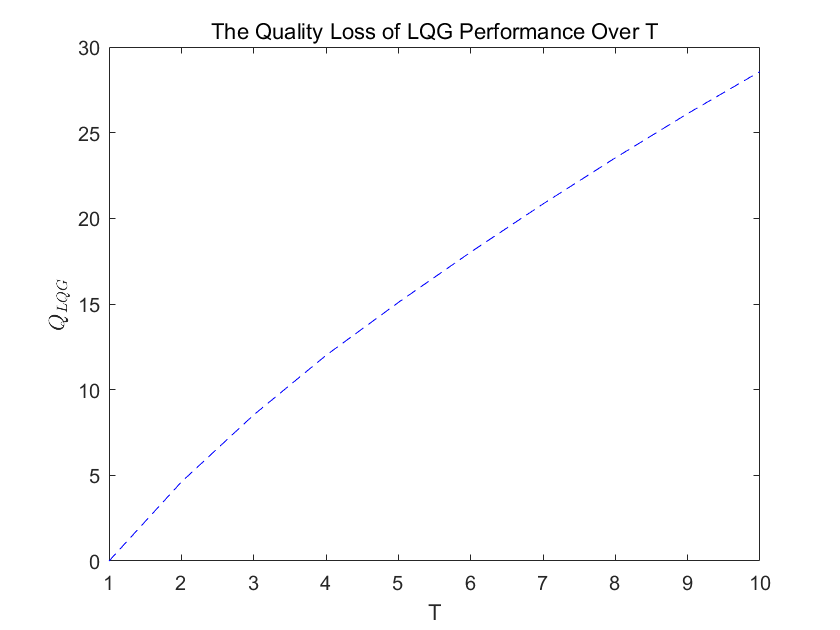}
		\caption{$\mathcal{Q}_{LQG}$ with $T$ varying from 1 to 10.}
		\label{LQGloss}
	\end{figure}

 
	Then we study the relationship between LQG performance loss $\mathcal{Q}_{LQG}$ and transmission period $T$ (Fig. \ref{LQGloss}).
	From the Fig. \ref{LQGloss}, $\mathcal{Q}_{LQG}$ is also monotonically increasing, which means the less information transmitted, the larger the cost of LQG control is. Recalling Theorem 2, we can find $\mathcal{Q}_{LQG}$ is linear to $\mathcal{Q}_{privacy}$. Hence their curves are similar. \par
\begin{figure}[htbp!]
		\centering
		\includegraphics[scale=0.33]{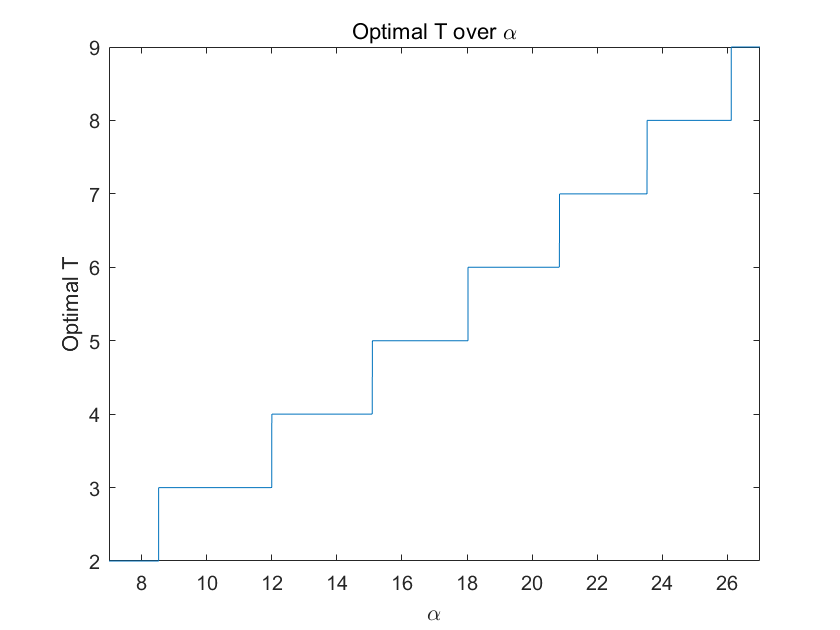}
		\caption{Optimal $T$ with $\alpha$ varying from 7 to 27.}
		\label{OptimalT}
	\end{figure}

 
	Given different thresholds of LQG performance loss $\alpha$, we can get optimal $T$. We plot optimal $T$ when $\alpha$ varies from 7 to 27 (Fig. \ref{OptimalT}).
	We can find that the curve is stepped, and it means for a continuous range of $\alpha$ the optimal $T$ is the same because it is discrete. As the given LQG performance loss threshold increases, a larger transmission period can be selected. \par
\begin{figure}[htbp!]
		\centering
		\includegraphics[scale=0.33]{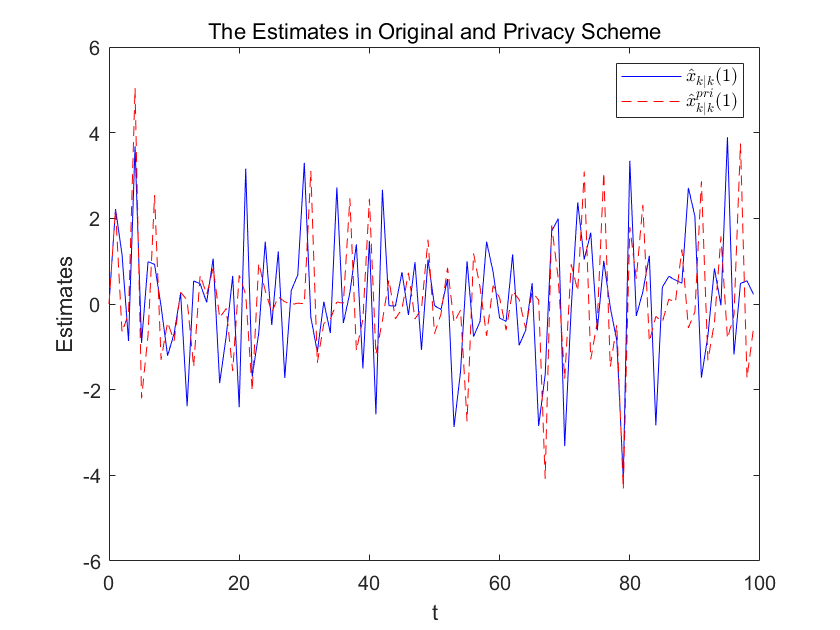}
		\caption{Estimates in two different schemes.}
		\label{Estimates}
	\end{figure}

 
	Finally we plot the estimates in two different schemes when $T=3$ (Fig. \ref{Estimates}), and we find that $\hat{x}_{k|k}^{pri}$ is deviated from $\hat{x}_{k|k}$.

	\section{Conclusion}
	In this paper, we consider a cooperative LQG control system with a single user and a server. We propose a scheme of intermittently transmission to preserve the user's privacy, and the corresponding privacy metric. Then we analyze the privacy level under the privacy scheme and the loss in LQG control performance. Finally, we propose an optimization problem and solve it based on the trade-off between the privacy level and the loss in LQG control performance. For the future work, the multi-sensor system and new privacy metric could be considered.
	
	\appendix
	For notional convenience, in the following sections, we omit the superscript ``\textit{pri}" in $\hat{x}_{k|k}^{pri}$, $P_{k|k}^{pri}$, $x_{k|k-1}^{pri}$, and $P_{k|k-1}^{pri}$ under the privacy scheme.  Furthermore, we use $\hat{x}_{k}$ and $P_{k}$ to replace $\hat{x}_{k|k}$ and $P_{k|k}$, and use $\hat{x}_{k}^{-}$ and $P_{k}^{-}$ to replace $\hat{x}_{k|k-1}$ and $P_{k|k-1}$.
	\subsection{Proof of Theorem 1}
		 To prove the theorem, we follow the procedure in Sec. 4.4 of \cite{anderson2012optimal}. We show that $P_{k}$ has two limits. When $k \to \infty$, the infimum limit corresponds to the covariance at the time the server receives $y_{k}$, i.e., $P_{lT+1}$ ($l \in \mathbb{Z}_{+}$). Meanwhile the supremum limit corresponds to the covariance at the previous time before the server receives $y_{k}$, i.e., $P_{lT}$ ($l \in \mathbb{N}$). Since if $P_{lT+1}^{-}$ converges, $P_{lT+1}=\tilde{g}(P_{lT+1}^{-})$ also converges. Hence, we focus on the sequence of $\left\{P_{lT+1}^{-}\right\}$. For general $k$, we firstly have 
		 \begin{align*}
		 	\hat{x}_{k}=\;& \hat{x}^{-}_{k}+K_{k}(y_{k}-C\hat{x}_{k}^{-}) \\
		 	=\;& (I-K_{k}C)\hat{x}_{k}^{-}+K_{k}Cx_{k}+K_{k}v_{k}.
		 \end{align*}
	 	Then 
	 	\begin{align*}
	 		x_{k}-\hat{x}_{k}=(I-K_{k}C)(x_{k}-\hat{x}^{-}_{k})-K_{k}v_{k}.
	 	\end{align*}
	 	We have 
	 	\begin{align*}
	 		P_{k}=(I-K_{k}C)P_{k}^{-}(I-K_{k}C)'+K_{k}RK_{k}'.
	 	\end{align*}
	 	Hence, when $k=(l-1)T+1$, we obtain
	 	\begin{align*}
	 		P_{(l-1)T+1}=\;&(I-K_{(l-1)T+1}C)P_{(l-1)T+1}^{-}(I-K_{(l-1)T+1}C)'\\
	 		\;&+K_{(l-1)T+1}RK_{(l-1)T+1}'.
	 	\end{align*}
 		 Since when the estimator does not receive $y_{k}$, the calculation of estimate error covariance follows eqn. (21), and we have
 		 \begin{align*}
 		 	P_{lT}=h^{T-1}(P_{(l-1)T+1}).
 		 \end{align*}
 	 	 Then  
 	 	 \begin{align*}
 	 	 	P_{lT+1}^{-} 
 	 	 	=\;& h(P_{lT}) \\
 	 	 	=\;&A^{T}(I-K_{(l-1)T+1}C)P_{(l-1)T+1}^{-}(I-K_{(l-1)T+1}C)'(A')^{T} \\
 	 	 	\;&+A^{T}K_{(l-1)T+1}RK_{(l-1)T+1}'(A')^{T}+\sum_{i=0}^{T-1}A^{i}Q(A')^{i}.
 	 	 \end{align*}
  	 	 We present the following steps to complete the proof. \\ \\ 
  	 	 \textbf{Step 1} \par  
	 	 Firstly we prove that $P_{lT+1}^{-}$ is bounded. Since we assume $(C,A)$ is detectable, there exits a $\tilde{K}$ which can formulate a suboptimal, asymptotically stable filter by 
	 	\begin{align*}
	 		\tilde{x}_{k}^{-}=A\tilde{x}_{k-1}^{-}+A\tilde{K}(y_{k}-C\tilde{x}_{k}^{-})+Bu_{k-1},
	 	\end{align*} 
 		and its corresponding error covariance is 
 		\begin{align*}
 			\tilde{P}_{k}^{-}=A(I-\tilde{K}C)\tilde{P}_{k-1}^{-}(I-\tilde{K}C)'A'+A\tilde{K}R\tilde{K}'A'+Q.
 		\end{align*}
 		Hence, we can obtain
 		\begin{align*}
 			\tilde{P}_{lT+1}^{-}  
 			=\;&A^{T}(I-\tilde{K}C)\tilde{P}_{(l-1)T+1}^{-}(I-\tilde{K}C)'(A')^{T} \\
 			\;&+A^{T}\tilde{K}R\tilde{K}'(A')^{T}+\sum_{i=0}^{T-1}A^{i}Q(A')^{i}.
 		\end{align*}
 		We assume $(C,A^{T})$ is detectable, which makes  $\rho(A^{T}(I-\tilde{K}C))<1$, and $A^{T}(I-\tilde{K}C)$ is stable and a constant matrix. Therefore, the result has a bounded solution. Based on the optimality of Kalman filter, we have $P_{lT+1}^{-}\leq \tilde{P}_{lT+1}^{-}$, hence $P_{lT+1}^{-}$ is also bounded. \\ \\
 		\textbf{Step 2} \par  
 		Next we prove the incrementality of $P_{lT+1}^{-}$. We firstly assume $P_{1}^{-}=0$. By simple computation, it can be obtained that $P_{T+1}^{-}\geq P_{1}^{-}=0$. For general $l$, 
 		\begin{align*}
 			&P_{lT+1}^{-} \\
 			=\;&A^{T}\!(I\!-\!K_{(l-1)T+1}C)P_{(l-1)T+1}^{-}(I\!-\!K_{(l-1)T+1}C)'\!(A')^{T} \\
 			\;&+A^{T}K_{(l-1)T+1}RK_{(l-1)T+1}'(A')^{T}+\sum_{i=0}^{T-1}A^{i}Q(A')^{i} \\
 			=\;&\min_{K}[A^{T}(I-KC)P_{(l-1)T+1}^{-}(I-KC)'(A')^{T} \\
 			\;&+A^{T}KRK'(A')^{T}+\sum_{i=0}^{T-1}A^{i}Q(A')^{i}] \\
 			\leq\;&A^{T}(I-K_{lT+1}C)P_{(l-1)T+1}^{-}(I-K_{lT+1}C)'(A')^{T} \\
 			\;&+A^{T}K_{lT+1}RK_{lT+1}'(A')^{T}+\sum_{i=0}^{T-1}A^{i}Q(A')^{i} \\
 			\leq\;& A^{T}(I-K_{lT+1}C)P_{lT+1}^{-}(I-K_{lT+1}C)'(A')^{T} \\
 			\;&+A^{T}K_{lT+1}RK_{lT+1}'(A')^{T}+\sum_{i=0}^{T-1}A^{i}Q(A')^{i} \\
 			=\;&P_{(l+1)T+1}^{-}.
 		\end{align*}
 	Now we have proved the incrementality and boundedness of $P_{lT+1}^{-}$, hence $P_{lT+1}^{-}$ is asymptotically stable, and so is $P_{lT+1}$. For $P_{lT+1}^{-}$ we have
 	\begin{align*}
 		P_{lT+1}^{-}=h^{T} \circ \tilde{g}(P_{(l-1)T+1}^{-}).
 	\end{align*}
 	Denote $\bar{\Sigma}$ is the asymptotically stable value of $P_{lT+1}^{-}$, and it is the unique solution to the following equation:
 	\begin{align*}
 		\bar{\Sigma}=h^{T} \circ \tilde{g}(\bar{\Sigma}).
 	\end{align*}
 	 We denote the infimum value of $P_{k}$ as $\tilde{P}$, and it is simple to see  that $\tilde{P}$ is the asymptotically stable value of $P_{lT+1}$, hence we have 
 	\begin{align*}
 		\tilde{P}=\tilde{g}(\bar{\Sigma})=\tilde{g} \circ h^{T} \circ \tilde{g}(\bar{\Sigma})=g \circ h^{T-1}(\tilde{P}).
 	\end{align*}
 	Therefore, we obtain the following results:
 	\begin{align*}
 		\lim_{l \to \infty}P_{lT+1}=\tilde{P},
 	\end{align*}
 	where $\tilde{P}$ is the unique solution to follows:
 	\begin{align*}
 		\tilde{P}=g \circ h^{T-1}(\tilde{P}).
 	\end{align*} 
 	Meanwhile, two limits of $P_{k}$ are given as
 	\begin{align*}
 		\liminf_{k \to \infty}P_{k}&=\tilde{P}, \\
 		\limsup_{k\to\infty}P_{k}&=h^{T-1}(\tilde{P}).
 	\end{align*}
 	In summary, we prove Theorem 1 under the zero initial covariance condition. \\ \\ 
 	\textbf{Step 3} \par   
	Finally we consider $P_{1}^{-}\neq 0$. Define a transition matrix $\Psi_{l}$ as 
	\begin{align*}
		\Psi_{l}=\;&[A^{T}(I-K_{(l-1)T+1}C)][A^{T}(I-K_{(l-2)T+1}C)] \\
		  \;&[A^{T}(I-K_{(l-3)T+1}C)] \cdots[A^{T}(I-K_{1}C)].
	\end{align*}
	Then $P_{lT+1}^{-}$ follows 
	\begin{align*}
		P_{lT+1}^{-}&=\Psi_{l}P_{1}^{-}\Psi_{l}'+nonnegative \enspace definite \enspace terms \\
		&\geq \Psi_{l}P_{1}^{-}\Psi_{l}'.
	\end{align*}
	We firstly take $P_{1}^{-}=\beta I (\beta >0)$, and it follows that $\beta \Psi_{l}\Psi_{l}'$. Since $P_{lT+1}^{-}$ is proved to be bounded before, $\Psi_{l}$ is also bounded. Assume that $P_{lT+1}^{-}$ has a asymptotically stable value at this scenario. We rewrite $P_{lT+1}^{-}$ as follows:
	\begin{align*}
		P_{lT+1}^{-}=h^{T} \circ \tilde{g}(P_{(l-1)T+1}^{-})=h^{T-1} \circ \big(h \circ \tilde{g}(P_{(l-1)T+1}^{-})\big).
	\end{align*}
	Hence, we have
	\begin{align*}
		&P_{lT+1}^{-} \\
		=\;&h^{T-1}(A(I-K_{(l-1)T+1}C)P_{(l-1)T+1}^{-}A'+Q) \\
		=\;&h^{T-1}(AP_{(l-1)T+1}^{-}(I-K_{(l-1)T+1}C)^{'}A'+Q) \\
		=\;&A^{T}P_{(l-1)T+1}^{-}(I-K_{(l-1)T+1}C)'(A')^{T}+\sum_{i=0}^{T-1}A^{i}Q(A')^{i}. 
	\end{align*}
	And $\bar{\Sigma}$ can also be computed as follows:
	\begin{align*}
		\bar{\Sigma} 
		=\;&h^{T-1}(A(I-KC)\bar{\Sigma}A'+Q) \\
		=\;&A^{T}(I-KC)\bar{\Sigma}(A')^{T}+\sum_{i=0}^{T-1}A^{i}Q(A')^{i}.
	\end{align*}
	By subtraction, it is follows that
	\begin{align*}
		&P_{lT+1}^{-}-\bar{\Sigma} \\
		=\;&A^{T}(I\!-\!KC)(P_{(l-1)T+1}^{-}\!-\!\bar{\Sigma})(I\!-\!K_{(l-1)T+1}C)'(A')^{T} \\
		\;&+A^{T}KCP_{(l-1)T+1}^{-}(I-K_{(l-1)T+1}C)'(A')^{T} \\
		\;&-A^{T}(I-KC)\bar{\Sigma}C'K_{(l-1)T+1}'(A')^{T}.
	\end{align*}
	For the last two terms, we have 
	\begin{align*}
		&A^{T}KCP_{(l-1)T+1}^{-}(I-K_{(l-1)T+1}C)'(A')^{T} \\
		&-A^{T}(I-KC)\bar{\Sigma}C'K_{(l-1)T+1}'(A')^{T} \\
		=\;&A^{T}KCP_{(l-1)T+1}(A')^{T}-A^{T}\tilde{P}C'K_{(l-1)T+1}'(A')^{T}.
	\end{align*}
	Substitute $K_{k}=P_{k}C'R^{-1}$ to it, and we obtain the result is zero.
	Hence, we get 
	\begin{align*}
		&P_{lT+1}^{-}-\bar{\Sigma} \\
		=\;&A^{T}(I\!-\!KC)(P_{(l-1)T+1}^{-}\!-\!\bar{\Sigma})(I\!-\!K_{(l-1)T+1}C)'(A')^{T}. 
	\end{align*}
	Suppose $P_{T}=\beta I$, we have
	\begin{align*}
		P_{lT+1}^{-}-\bar{\Sigma} 
		=[A^{T}(I-KC)]^{l}(P_{1}^{-}-\bar{\Sigma})\Psi_{l}'.
	\end{align*}
	Due to $(C,A^{T})$ is detectable, and $\Psi_{l}$ is bounded, when $l \to \infty$, we have $P_{lT+1}^{-}=\bar{\Sigma}$. \par 
	For an arbitrary $P_{1}^{-}$, we take such $\beta>0$ that $\beta I>P_{1}^{-}$, and denote the solution to $X=h^{T} \circ \tilde{g}(X)$ as $\bar{\Sigma}_{\beta I-}$. Meanwhile we denote $\bar{\Sigma}_{0}$ and $\bar{\Sigma}_{\beta I}$ as the solution to the same equation when $P_{1}^{-}=0$ and $P_{1}^{-}=\beta I$. We can simply get
	\begin{align*}
		\bar{\Sigma}_{0}<\bar{\Sigma}_{\beta I-}<\bar{\Sigma}_{\beta I}.
	\end{align*}
	From the previous proof, both $\bar{\Sigma}_{0}$ and $\bar{\Sigma}_{\beta I}$ tend to $\bar{\Sigma}$, hence $\bar{\Sigma}_{\beta I-}$ converges to $\bar{\Sigma}$ as well, i.e., $P_{lT+1}^{-}$ converges at this scenario. Therefore, $P_{lT+1}$ also converges. Then we can obtain the same conclusions as in the case of zero initial covariance.

	\subsection{Proof of Theorem 2}
		We consider the case of transmission period $T=2$, and other cases can also be proved similarly. In this case, according to eqn. (14), for $l\in \mathbb{Z}_{+}$, the transmission scheme is formulated as follows:
		\begin{align*}
			\gamma_{k}=\begin{cases}
				1, & k=2l+1, \\
				0, & k \neq 2l+1. 
			\end{cases} 
		\end{align*}  
		Firstly we consider the scenario of a finite-time horizon $N$, and assume that $N=2L+1$, where $L$ is a positive integer. Then at time $k=N-1$ the measurement is not sent, and at time $k=N-2$ it is sent.  Define the information set $I_k$ as
		\begin{align*}
			I_{0}=\;& \left\{\emptyset\right\}, \\
			I_{k}=\;& \left\{\gamma_{1}y_{1}, \dots, \gamma_{k}y_{k},u_{0}, \dots, u_{k-1}\right\}, \quad k=1, \dots , N.
		\end{align*}
		The finite-time quadratic objective
		function for the LQG control is denoted as $J_{0:N}$, and is defined as follows:
		\begin{align*}
			J_{0:N}\triangleq  \mathbf{E}\Big[\sum_{k=0}^{N-1}(x_{k}'Wx_{k}+u_{k}'Uu_{k})+x_{N}'Wx_{N}\Big].
		\end{align*}
		Meanwhile, we define $J_{k:N}$ as 
		\begin{align*}
			J_{k:N}=\min_{u_{k}}\mathbf{E}\Big[x_{k}'Wx_{k}+u_{k}'Uu_{k}+J_{k+1} |I_{k}\Big].
		\end{align*}
		At time $N$, we have $J_{N:N}=\mathbf{E}[x_N'W{x_N}|I_{N}]$. Therefore, at time $k=N-1$, 
		\begin{align*}
			&J_{N-1:N} \\
			\;\;=&\min_{u_{N-1}}\mathbf{E}\Big[x_{N-1}'Wx_{N-1}+u_{N-1}'Uu_{N-1}+J_{N:N}|I_{N-1}\Big]  \\
			\;\;=&\min_{u_{N-1}}\mathbf{E}\Big[x_{N-1}'Wx_{N-1}+u_{N-1}'Uu_{N-1}+x_{N}'Wx_{N}|I_{N-1}\Big] \\
			=&\min_{u_{N-1}}\mathbf{E}\Big[u_{N-1}'Uu_{N-1}+x_{N-1}'A'WBu_{N-1} \\
			&+u_{N-1}'B'WAx_{N-1}+u_{N-1}'B'WBu_{N-1}|I_{N-1}\Big] \\
			&+\mathbf{E}\Big[x_{N-1}'Wx_{N-1}+x_{N-1}'A'WAx_{N-1}|I_{N-1}\Big]+tr(WQ). 
		\end{align*} 
		Since the measurement is not sent at this time, for the server we have $\hat{x}_{N-1}=\hat{x}_{N-1}^{-}$, i.e., $\mathbf{E}(x_{N-1}|I_{N-1})=\hat{x}_{N-1}=\hat{x}_{N-1}^{-}$. Hence, we have
		\begin{align*}
			&J_{N-1:N} \\
			=&\min_{u_{N-1}}\Big[u_{N-1}'(U+B'WB)u_{N-1}+2u_{N-1}'B'WA\hat{x}_{N-1}^{-}\Big] \\		&+\mathbf{E}\Big[x_{N-1}'(W+A'WA)x_{N-1}|I_{N-1}\Big]+tr(WQ).
		\end{align*} 
		To find the minimum value, we take the derivative of the term containing $u_{N-1}$ in the above result and let it be zero:
		\begin{align*}
			2(B'WB+U)u_{N-1}^{*}+2B'WA\hat{x}_{N-1}^{-}=0,
		\end{align*}
		which leads to
		\begin{align*}
			u_{N-1}^{*}=-(B^{'}WB+U)^{-1}B^{'}WA\hat{x}_{N-1}^{-}.
		\end{align*}
		Substituting it into the above equation:
		\begin{align*}
			&J_{N-1:N} \\
			=\;&\mathbf{E}\Big[x_{N-1}'(W+A'WA)x_{N-1}|I_{N-1}\Big]+tr(WQ) \\
			\;&-(\hat{x}^{-}_{N-1})'A'WB(B'WB+U)^{-1}B'WA\hat{x}_{N-1}^{-}.
		\end{align*}
		Let 
		\begin{align*}
			S_{N}&=W, \\
			\Phi_{N-1}&=A'S_{N}B\left(B'S_{N}B+U\right)^{-1}B'S_{N}A, \\
			S_{N-1}&=A'S_{N}A+W-\Phi_{N-1}.
		\end{align*}
		Then we have 
		\begin{align*}
			&J_{N-1:N} \\
			=\;&\mathbf{E}\Big[x_{N-1}'S_{N-1}x_{N-1}|I_{N-1}\Big]+tr(S_{N}Q) \\
			\;&+\mathbf{E}\Big[x_{N-1}'\Phi_{N-1}x_{N-1}|I_{N-1}\Big]-(\hat{x}_{N-1}^{-})'\Phi_{N-1}\hat{x}_{N-1}^{-} \\
			=\;&\mathbf{E}\!\Big[x_{N-1}'S_{N-1}x_{N-1}|I_{N-1}\Big]\!+\!tr\!(S_{N}Q)\!+\!tr\!(\Phi_{N-1}\!P_{N-1}^{-}).
		\end{align*}
		At time $k=N-2$,
		\begin{align*}
			&J_{N-2:N} \\
			=&\min_{u_{N-2}}\mathbf{E}\Big[x_{N-2}'Wx_{N-2}\!+\!u_{N-2}'Uu_{N-2}
			\!+\!J_{N-1:N}|I_{N-2}\Big]  \\
			=&\min_{u_{N-2}}\Big\{u_{N-2}'Uu_{N-2}+\mathbf{E}\left(x_{N-1}'S_{N-1}x_{N-1}|I_{N-2}\right)\Big\} \\
			&\!+\!\mathbf{E}\Big[x_{N-2}'Wx_{N-2}|I_{N-2}\!\Big]\!+\!tr(S_{N}Q)\!+\!tr\!(\Phi_{N-1}P_{N-1}^{-}\!) \\
			=&\min_{u_{N-2}}\Big\{u_{N-2}'Uu_{N-2}+\mathbf{E}\Big[(Ax_{N-2}+Bu_{N-2}+w_{N-2})'\\
			&S_{N-1}(Ax_{N-2}+Bu_{N-2}+w_{N-2})|I_{N-2}\Big]\Big\} \\
			&\!+\!\mathbf{E}\!\Big[x_{N-2}'Wx_{N-2}|I_{N-2}\Big]\!+\!tr(S_{N}Q)\!+\!tr(\Phi_{N-1}P_{N-1}^{-}) \\
			=&\min_{u_{N-2}}\Big\{u_{N-2}'(U+B'S_{N-1}B)u_{N-2} \\
			&\!+\!\!\mathbf{E}\!\Big[x_{N-2}'A'\!S_{N-1}Bu_{N-2}\!+\!u_{N-2}'B'\!S_{N-1}Ax_{N-2}|I_{N-2}\Big]\!\Big\} \\
			&\!+\mathbf{E}\Big[x_{N-2}'A'S_{N-1}Ax_{N-2}|I_{N-2}\Big]+tr(S_{N-1}Q) \\
			&\!+\!\mathbf{E}\Big[x_{N-2}'Wx_{N-2}|I_{N-2}\Big]\!+\!tr\!(S_{N}Q)\!+\!tr\!(\Phi_{N-1}P_{N-1}^{-}) \\
			=&\min_{u_{N-2}}\Big\{u_{N-2}'(U+B'S_{N-1}B)u_{N-2} \\
			&\!+\!\!\mathbf{E}\!\Big[x_{N-2}'A'\!S_{N-1}Bu_{N-2}\!+\!u_{N-2}'B'\!S_{N-1}Ax_{N-2}|I_{N-2}\Big]\Big\} \\
			&\!+\mathbf{E}\Big[x_{N-2}'(A'S_{N-1}A+W)x_{N-2}|I_{N-2}\Big] \\
			&\!+tr\Big[(S_{N-1}+S_{N})Q\Big]+tr(\Phi_{N-1}P_{N-1}^{-}).
		\end{align*}
		Similarly, we take the derivative of the term containing $u_{N-2}$ in the above result and let it be zero:
		\begin{align*}
			2(U+B'S_{N-1}B)u_{N-2}^{*}+2B'S_{N-1}A\hat{x}_{N-2}=0,
		\end{align*}
		which leads to
		\begin{align*}
			u_{N-2}^{*}=-(U+B'S_{N-1}B)^{-1}B'S_{N-1}A\hat{x}_{N-2}.
		\end{align*}
		Substituting it into the above equation, we obtain
		\begin{align*}
			&J_{N-2:N} \\
			=\;&\mathbf{E}\Big[x_{N-2}'(A'S_{N-1}A+W)x_{N-2}|I_{N-2}\Big] \\
			\;&+tr\Big[(S_{N-1}+S_{N})Q\Big]+tr(\Phi_{N-1}P_{N-1}^{-}) \\
			\;&-\!x_{N-2}'A'\!S_{N-1}B\!(U\!+\!B'S_{N-1}B)\!^{-1}B'\!S_{N-1}A\hat{x}_{N-2}.
		\end{align*}
		Let 
		\begin{align*}
			\Phi_{N-2}&=A^{'}S_{N-1}B(U+B^{'}S_{N-1}B)^{-1}B^{'}S_{N-1}A, \\
			S_{N-2}&=A^{'}S_{N-1}A+W-\Phi_{N-2}. 
		\end{align*}
		Then 
		\begin{align*}
			&J_{N-2:N} \\
			=\;&\mathbf{E}\Big[x_{N-2}'S_{N-2}x_{N-2}|I_{N-2}\Big]+\mathbf{E}\Big[x_{N-2}'\Phi_{N-2}x_{N-2}|I_{N-2}\Big]\\
			\;&-\hat{x}_{N-2}'\Phi_{N-2}\hat{x}_{N-2}+tr\Big[(S_{N-1}+S_{N})Q\Big] +tr(\Phi_{N-1}P_{N-1}^{-}) \\
			=\;&\mathbf{E}\Big[x_{N-2}'S_{N-2}x_{N-2}|I_{N-2}\Big]+tr\Big[(S_{N-1}+S_{N})Q\Big] \\
			\;&+tr(\Phi_{N-1}P_{N-1}^{-})+tr(\Phi_{N-2}P_{N-2}).
		\end{align*}
		Similarly, for any time $k$, we have
		\begin{align*}
			\Phi_{k}&=A'S_{k+1}B(U+B'S_{k+1}B)^{-1}B'S_{k+1}A, \\
			S_{N}&=W,\\
			S_{k}&=A'S_{k+1}A+W-\Phi_{k}, \\
			L_{k}&=-(B'S_{k+1}B+U)^{-1}B'S_{k+1}A,\\
			u_{k}^{*}&=L_{k}\mathbf{E}[x_{k}|I_{k}].
		\end{align*}
		Then the optimal objective function at time $k$ is
		\begin{align*}
			J_{k:N}=\mathbf{E}(x_{k}'S_{k}x_{k}|I_{k})+r_{k}+t_{k},
		\end{align*}
		where
		\begin{align*}
			r_{N}&=0,\\
			r_{k}&=r_{k+1}+tr(S_{k+1}Q),\\
			t_{N}&=0,\\
			t_{k}&=\begin{cases}
				t_{k+1}+tr(\Phi_{k}P_{k}^{-}), & k\neq 2l+1, \\
				t_{k+1}+tr(\Phi_{k}P_{k}), & k=2l+1.
			\end{cases} 
		\end{align*}
		Hence, the optimal objective function is
		\begin{align*}
			J_{0:N} 
			=\;&\mathbf{E}(x_{0}^{'}S_{0}x_{0})+\sum_{k=0}^{N-1}tr(S_{k+1}Q) \\
			\;&+\!\!\!\!\!\!\!\!\!\!\!\!\!\!\! \sum_{k\neq 2l+1,\; k \in [0,N-1]}\!\!\!\!\!\!\!\!\!\!\!\! tr(\Phi_{k}P_{k}^{-}) +\!\!\!\!\!\!\!\!\!\!\!\!\!\!\! \sum_{k=2l+1,\; k \in [0,N-1]}\!\!\!\!\!\!\!\!\!\!\!\! tr(\Phi_{k}P_{k}).
		\end{align*}
		After straightforward calculation we can see that, for other values of $T$, the similar conclusions can also be drawn through the similar proof steps. For general $T$, we have 
		\begin{align*}
			J_{0:N}
			=\;&\mathbf{E}(x_{0}^{'}S_{0}x_{0})+\sum_{k=0}^{N-1}tr(S_{k+1}Q) \\
			\;&+\!\!\!\!\!\!\!\!\!\!\!\!\!\!\! \sum_{k\neq lT+1,\; k \in [0,N-1]}\!\!\!\!\!\!\!\!\!\!\!\! tr(\Phi_{k}P_{k}^{-}) +\!\!\!\!\!\!\!\!\!\!\!\!\!\!\! \sum_{k=lT+1,\; k \in [0,N-1]}\!\!\!\!\!\!\!\!\!\!\!\! tr(\Phi_{k}P_{k}).
		\end{align*}
		Under the infinite-time horizon, we have
		\begin{align*}
			\mathcal{O}^{*}=&\lim_{N\to \infty} \frac{1}{N} J_{0:N}. 
		\end{align*}
		According to eqn. (26) and the above equations, we can write $\mathcal{O}^{*}$ as
		\begin{align*}
			\mathcal{O}^{*}&= tr(SQ)+tr\big(\Phi  \frac{1}{T}\sum_{i=0}^{T-1} h^{i}(\tilde{P})\big). 
		\end{align*}
	%
	%
		%
		%
	
	\section*{Acknowledgment}
	The work by Wen Yang and Chao Yang is supported by the National Natural Science Foundation of China under Grant 62336005. \par 
	The work by Yuqing Ni is supported by the National Natural Science Foundation of China under Grant 62303196, the Natural Science Foundation of Jiangsu Province of China under Grant BK20231036, and the Basic Research Funds of Wuxi Taihu Light Project under Grant K20221005.
	
	\bibliographystyle{IEEEtran}
	\bibliography{Privacy_Intermittent_IEEEtran}
	
\end{document}